\documentstyle[aps,epsf]{revtex}  % DO NOT DELETE THIS LINE 

\begin{document}        %  DO NOT DELETE OR CHANGE THIS LINE

\baselineskip 14pt
\title{First Charm Baryon Physics from SELEX(E781)}
\author{Fernanda G. Garcia}
\address{Departamento de F\'{\i}sica, Universidade de S\~ao Paulo, USP,
         S\~ao Paulo, S.P., Brasil}
\author{Soon Yung Jun} % Use this and the next line only if there is a second
\address{Carnegie Mellon University, Pittsburgh, PA 15213}  % address.
\author{\it (For the SELEX Collaboration~\cite{SELEX})} % Use this and the next line only if there is a second
\address{}  % address.
\maketitle              % Creates the title area, Do Not Remove

\begin{abstract}        % Do Not Delete this line
We present preliminary results on various aspects of charm baryon studies
at the 1996-1997 fixed target experiment of Fermilab studying charm
produced from incident $\Sigma^-$, proton, and $\pi^-$ beams at 600 GeV.
First results include the comparison of hadroproduction asymmetries for 
$\Lambda_c^+$ production from the 3 beams as well $x_F$ distributions and the
first observation of the Cabbibo-suppressed decay 
$\Xi_c^+ \rightarrow p K^- \pi^+$.  The relative branching fraction of the 
Cabbibo-suppressed mode to the 3-body Cabbibo-favored modes is also
presented.
\
\end{abstract}   	% Do Not Delete this line

\section{Introduction}               % Introduction goes below.

The hadroproduction of the particles with open charm is a rich area to explore
the hadronization process of heavy quarks. In hadronic interactions, it is
especially interesting to compare the forward production characteristics of 
a given charm particle species for different beam hadrons.  The hadronization
process may introduce a distinction between those charm hadrons which have a 
valence quark or anti-quark in common with the beam compared to those which 
do not.  We assume that charm production can be factorized 
into perturbative and non-perturbative elements. The first part describes the 
production of the pair $c \overline{c}$ which the $gg$ fusion mechanism is the
main diagram. This is calculable in perturbative QCD (pQCD), which
does not predict any asymmetry between the $c$ and the $\overline c$ 
produced. Next--to--leading order pQCD calculations introduce a small 
asymmetry in the quark momenta~\cite{Nason,Frixione}. 
The soft part, or hadronization, describes the process by which charmed
quarks appear as hadrons in the final state. Only phenomenological 
models exist at this stage.  For $\pi^-$ beams a large asymmetry has been 
observed experimentally~\cite{E769,E791}.  We present here preliminary 
results for charm-anticharm 
asymmetries for $\Lambda_c^+$ production from three different beams.

Charm decays are also important laboratories for understanding pQCD at
the charm mass scale.
State--of--the--art methods for calculating non-leptonic decay rates of the 
charmed hadrons employ heavy quark effective theory and the factorization
approximation~\cite{HQEH}.  Nonetheless, the three--body Cabibbo--suppressed 
decays of charmed baryons are prohibitively difficult to calculate.  
Measurements of relative branching fraction  
between simple charmed baryon states, both Cabibbo--favored and 
Cabibbo--suppressed, give additional insight into the structure 
of the decay amplitude and the validity of the factorization approximation.  
Until now the only Cabibbo--suppressed charmed baryon 
decay reported with significant statistics is
$\Lambda_{c}^{+} \rightarrow pK^{-}K^{+}$ and its resonant 
state, $\Lambda_{c}^{+} \rightarrow p\phi$ ~\cite{CLEO-1,E687-1}.  
In this paper, we report the first observation of 
the Cabibbo--suppressed $\Xi_{c}^{+} \rightarrow pK^{-}\pi^{+}$ and determine 
the branching fraction of this decay relative to the Cabibbo--favored
$\Xi_{c}^{+} \rightarrow \Sigma^{+}K^{-}\pi^{+}$ mode.

\section{The SELEX Experiment}

SELEX is a high energy hadroproduction experiment at Fermilab using
a multi--stage spectrometer designed for high acceptance for forward 
interactions ($x_F = 2p_{\parallel}/\sqrt{s} > 0.1$).   The experiment 
aimed especially to understand charm production in the forward hemisphere 
and to study charmed baryon decays.  Using both a negative hyperon 
beam (50\% $\Sigma^-$, 50\% $\pi^-$) and a proton beam (92\% p, 8\% $\pi^+$),
SELEX took 15.2 billion interaction events
during the 1996--1997 fix target run, tagging 600 GeV $\Sigma^-$, $\pi^-$ 
and 540 GeV $p$ beams with a beam TRD.  The data were accumulated using a 
segmented charm target (5 foils: 2 Cu, 3 C, each separated by 1.5 cm) whose 
total thickness was  5\% of interaction length for  protons.

The spectrometer had silicon strip detectors to measure beam and outgoing
tracks and provided precision primary and secondary vertex reconstruction.
Transverse position resolution was 4 $\mu{\rm m}$ for the 600 GeV beam and 
the average longitudinal vertex position resolution was 
270 $\mu{\rm m}$ for primary and 560 $\mu{\rm m}$  for secondaries, 
respectively.  Track momenta were measured after magnetic deflection by
a system of proportional wire 
chambers (PWC), drift chambers and silicon strip detectors.  
Track momentum resolution for a typical 100 GeV track 
was $\Delta p/p \approx 0.5$\%.  The absolute
momentum scale was calibrated to the $K_s^0$ mass.  
For $D^0 \rightarrow K^- \pi^+$ decays the average mass resolution was 
9 MeV for $D^0$ momenta from 100-450 GeV.
Charged particle identification was done with the Ring Image 
$\check{\rm C}$erenkov Radiation (RICH) detector~\cite{rich-nim} which 
distinguished $K^{\pm}$ from $\pi^{\pm}$ up to 165 GeV.  The proton 
identification efficiency was greater than 95\% above proton 
threshold ($\approx 90$ GeV) and 
the pion mis--identification probability was about 4\%.  

Interactions were selected by a scintillator trigger and an online software
filter.  The charm trigger required at least 4 charged tracks and 
2 hits in a scintillator hodoscope after the second analyzing magnet.  
We triggered on about 1/3 of all inelastic interactions.
Based on downstream tracking with particle identification information,  
the online filter selected events that had evidence for a secondary vertex
among tracks completely reconstructed using the forward PWC spectrometer and 
the vertex silicon.  This filter reduced the 
data size by a factor of nearly 8 at a cost of about a factor of 2 in 
charm written to tape as normalized from
a study of unfiltered $K_s^0$ and $\Lambda^0$ decays.
Most of the charm loss came from selection
cuts that were independent of charm species or kinematic variables and
which improved the signal/noise in the final sample.

The charm events were selected by the following requirements;
(1) fits for both primary and secondary vertex have $\chi^{2}/dof <$ 5, 
(2) Longitudinal separation L between primary and secondary vertices is 
greater than 8 times the combined longitudinal error $\sigma$, and 
(3) the reconstructed momentum vector from the secondary vertex points 
back to the primary vertex with good quality $\chi^{2}$, and
(4) $\mathcal{L}({\rm K})/\mathcal{L}(\pi) > {\rm 1}$ for K identification and
$\mathcal{L}({\rm p})/\mathcal{L}(\pi) > {\rm 1}$ for $p$ identification, 
where $\mathcal{L}$ is the likelihood function based on RICH information.
Additional cuts will be explicitly described if applied.

\section{Charm Hadroproduction}

The production properties of charm quarks require measurements of charm 
hadrons.  pQCD calculations  
can be probed experimentally using measurements of single-charm-particle 
inclusive distributions as a function of 
$x_F$ and asymmetries, either integrated or as a function of $x_F$.

Previous charm hadroproduction experiments showed evidence 
of a large enhancement in the forward production of charmed particles that 
contain a quark or an antiquark in common with the beam ({\it leading 
particles}) over those that do not ({\it non-leading particles}),
in the meson sector. Recently, this study has been extended to the production
of baryons by a $\Sigma^-$ beam~\cite{WA89}. 

\begin{figure}
\centerline{\epsfxsize 4.2 truein \epsfbox{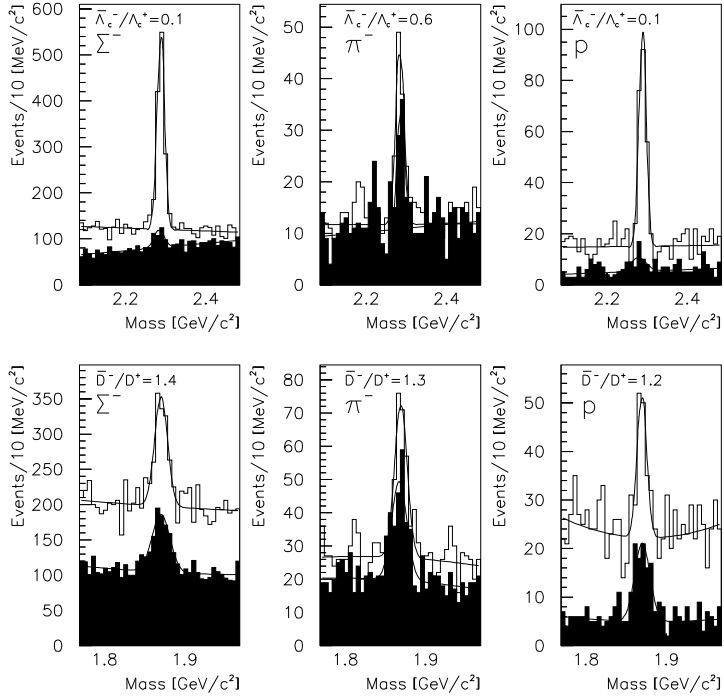}}   
\vskip -.2 cm
\caption[]{
\label{integrated_a}
\small $\Lambda_c^+ \rightarrow p K^- \pi^+$ and 
$D^+ \rightarrow K^- \pi^+ \pi^+$  signal for $x_F>0.3$ 
for $\Sigma^-$, $\pi^-$ and proton beams.}
\end{figure}

According to our simulations and our measurements of $K_s^0$ decays, the 
SELEX spectrometer acceptance is charge independent and we have a smooth, 
generally constant acceptance as a function of $x_F$ for $x_F>0.3$. We show in 
Fig.~\ref{integrated_a} the integrated charm production asymmetry 
for $\overline{\Lambda}_{c}^{-}/\Lambda_c^+$ and $D^-/D^+$ 
for $\Sigma^-$, $\pi^-$ and proton beams in the kinematic region $x_F>0.3$. 

\begin{figure}
\centerline{\epsfxsize 4.2 truein \epsfbox{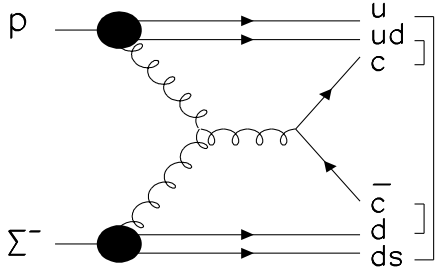}}   
\vskip -.2 cm
\caption[]{
\label{stFigure}
\small An example of $gg \rightarrow c{\overline c}$ fusion production 
mechanism in a p$\Sigma^-$ collision. The proton remnants are represented 
by a quark and a diquark.}
\end{figure}

The raw asymmetry for charm baryons is much stronger than for charm mesons
in this forward $x_F$ region for all 3 beam types.  The effect is even
more pronounced when the charm baryons are produced by a baryon beam.
One typical explanation of this observed asymmetry is that longitudinal 
momentum is added to the produced charm quark if it recombines with a 
valence quark from the incoming particle, forming the leading 
particle shown in Fig.~\ref{stFigure}.   This is incorporated, for example,
in the Pythia simulation program~\cite{lund}, where the effect is overestimated
for typical model parameters~\cite{E791}.
With this scenario in mind, we expect to find a harder $x_F$ spectrum for 
leading particles than non-leading particles. 
%The analysis of the
%differential production spectrum for each beam hadron is now underway.

%

\begin{figure}[ht]
\centerline{\epsfxsize 4.2 truein \epsfbox{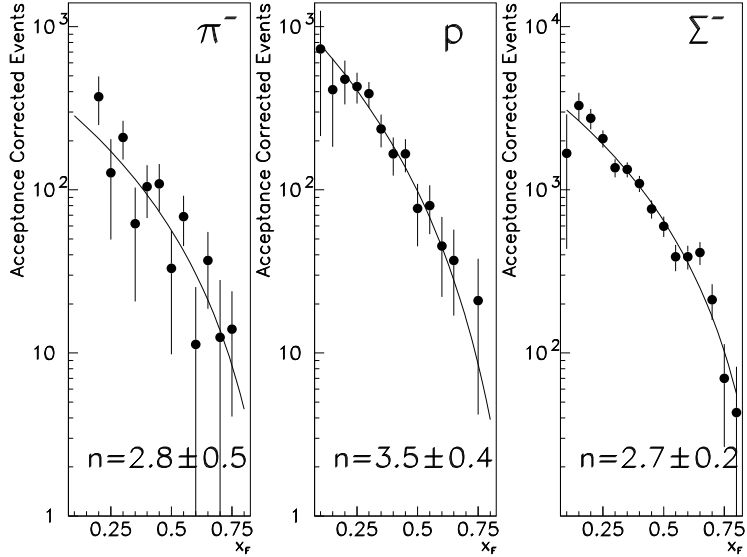}}   
\vskip -.2 cm
\caption[]{
\label{xf}
\small $\Lambda_c^+ \rightarrow p K^- \pi^+$ production as a function 
of $x_F$ for $\pi^-$, proton and $\Sigma^-$ beams.}
\end{figure}

The $x_F$ dependence of the corrected number of events from each beam is 
shown in figure~\ref{xf}. The curves are fits of the standard parametrization
$(1-x_F)^n$ to the data. 
The $x_F$ distributions for $\Lambda_c^+$ show an $x_F$ dependence for all 
three beam types that is harder than that reported for D mesons from a $\pi$ 
beam.~\cite{E791}. The $\Lambda_c^+$ is a leading particle for each of the 
three beam particles reported here, and the values of $n$ are all consistent.
Further insight into the mechanism will come from comparisons of charm
meson production characteristics from these three beam hadrons.  That work
is in process.

\section{Charm Baryon $\Xi_{c}^{+}$ Decays}

The study of Cabibbo--suppressed charm decays can provide useful insights into
the weak interaction mechanism for non-leptonic decays~\cite{Korner}.
The observed final state may arise either from direct quark mixing at the
decay stage or, in some cases, from quark rearrangement in final-state
scattering.  By comparing the strengths of Cabibbo--suppressed decays to their
Cabibbo--favored analogs, one can, in a systematic way, assess the 
contributions of the various mechanisms. 

\begin{figure}
\centerline{\epsfxsize 4.2 truein \epsfbox{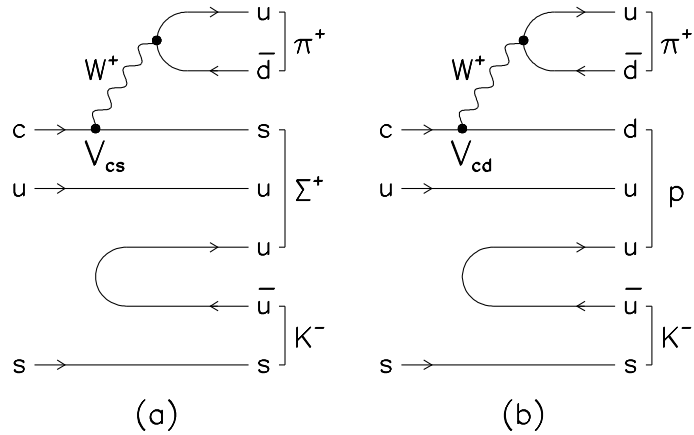}}   
\vskip -.2 cm
\caption[]{
\label{decayFigure}
\small An example of a spectator diagram with external $W$--emission 
for $\Xi_{c}^{+}$ Decays,
(a) Cabibbo--favored $\Sigma^{+}K^{-}\pi^{+}$ and 
(b) Cabibbo--suppressed  $pK^{-}\pi^{+}$.}
\end{figure}

Fig.~\ref{decayFigure} shows a simple spectator diagram with 
external $W$--emission for $\Xi_{c}^{+}$ decaying into  
a Cabibbo--allowed and a Cabibbo--suppressed mode.  The other Cabibbo--allowed
$\Xi^-$ mode interchanges $s$ and $d$ quark lines and produces a 
$d\overline{d}$ pair from the vacuum instead of a $d\overline {u}$ pair.

Since the decay process is similar in the two modes except flavor changing 
through the Cabibbo--Kobayashi--Maskawa matrix 
element ($V_{cs} {\rm vs.} V_{cd})$, we expect 
$B(\Xi_{c}^{+} \rightarrow pK^{-}\pi^{+})/
 B(\Xi_{c}^{+} \rightarrow \Sigma^{+}K^{-}\pi^{+}) = 
\alpha \times {\rm tan}^{2}\theta_{c}$,
where $\theta_{c}$ is the Cabibbo angle and $\alpha$ is a coefficient of 
order one containing information about differences in the two decay mechanisms 
over the allowed phase space.  To the extent that the relative branching 
ratio is different from this, we may argue for enhancement or suppression of
one of the two modes.

\begin{figure}
\centerline{\epsfxsize 4.2 truein \epsfbox{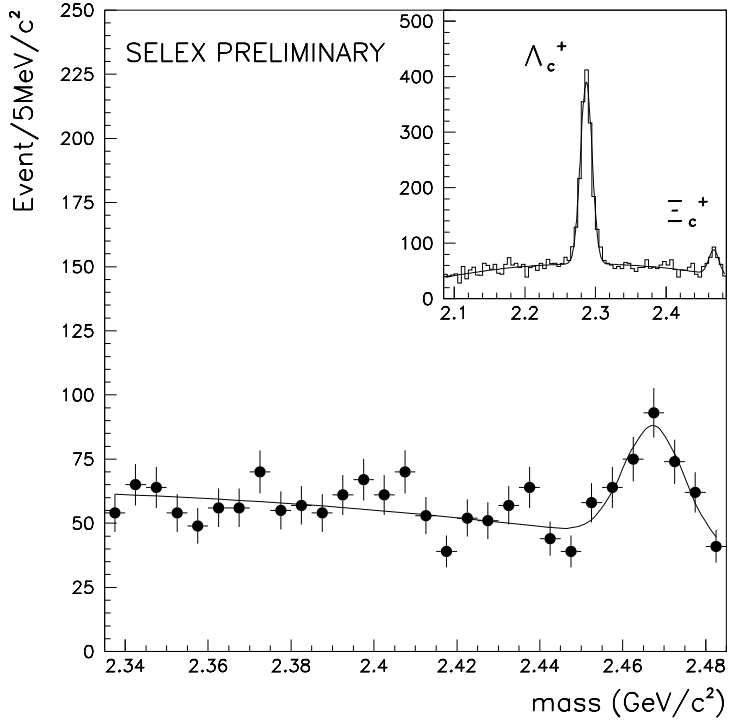}}   
\vskip -.2 cm
\caption[]{
\label{cspkiFigure}
\small The invariant mass distribution of $pK^-\pi^+$.}
\end{figure}

Fig.~\ref{cspkiFigure} shows the first observation of the Cabibbo--suppressed
$\Xi_{c}^+ \rightarrow p K^-\pi^+$ decay mode.
The inset of the figure shows the invariant mass distribution from 
reconstructed $p K^-\pi^+$ candidates with $p_{\pi} > 5$ GeV and
the sum of transverse momentum square ($\Sigma p_{T}^2$) of decaying 
particles greater than 0.3 (GeV$)^2$.  The large peak is the charm 
baryon decay, $\Lambda_{c}^+ \rightarrow pK^-\pi^+$ and the bump in the right 
corner shows the Cabibbo--suppressed $\Xi_{c}^+ \rightarrow pK^-\pi^+$ decay.
The analysis cuts off because of a maximum mass cut in the analysis of this
mode during this first pass through the data. 
We observed $162 \pm 31$ events for $\Xi_{c}^+ \rightarrow pK^-\pi^+$ with the 
$\Xi_{c}^+$ mass at ($2467.4 \pm 1.4$) MeV.  The statistical 
significance for the signal, $S/\sqrt{S+B}$, is $(7.0 \pm 1.3)$
in which $S$ is the number of signal events and $B$ is the number of 
background events under the signal region. 

\begin{figure}[ht]      % in second brace, h=here, t=top, b=bottom      
\centerline{\epsfxsize 4.2 truein \epsfbox{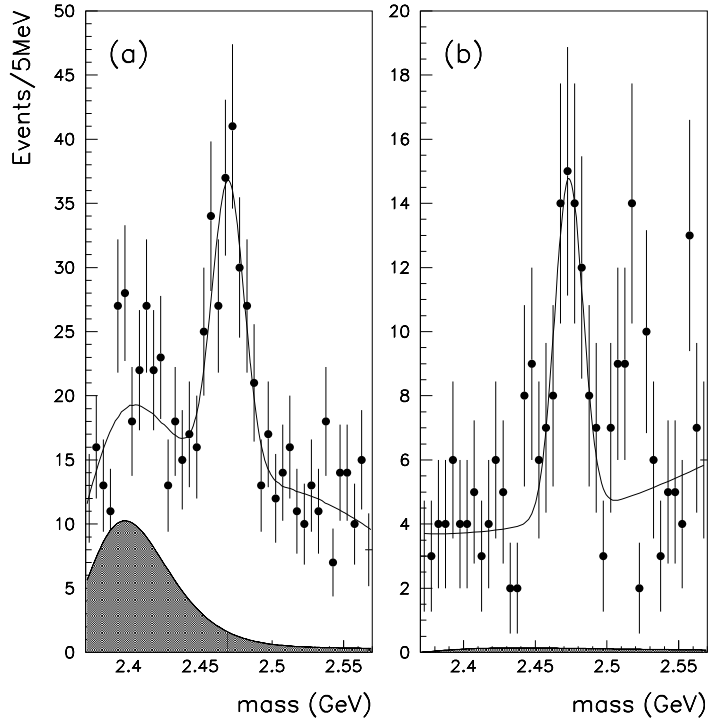}}   
%\vskip -.2 cm
\caption[]{
\label{brFigure}
\small Charmed baryon $\Xi_{c}^+$ signals with hyperon partial 
reconstruction, (a) $\Xi^-\pi^+\pi^+$ and (b) $\Sigma^+K^-\pi^-$.}
\end{figure}

In this first reconstruction pass, hyperons ($\Sigma^{\pm}, \Xi^{-}$)
were identified only inclusively in the limited decay interval 
($ 5{\rm m} < z_{decay} < 12 {\rm m})$.  The hyperon candidate track was 
identified with a track having $p > 40$ GeV for which no reconstructed
track segments were observed in the 14 chambers
along the trajectory after the second analyzing magnet. This category of
tracks (kinks by disappearance) had a unique $\Sigma^+$ identification for 
positive kinks but had an ambiguity for negative kinks between
$\Sigma^{-}$ and $\Xi^{-}$.   Reflections in 3-body modes are therefore
expected and are taken into account, based on data in the true mode.
Fig.~\ref{brFigure} shows two Cabibbo--favored $\Xi_{c}^+$ modes decaying to 
$\Xi^-\pi^+\pi^+$ and $\Sigma^+K^-\pi^-$, respectively.  For these modes, we 
require additional kinematical cuts: (a) the transverse component of the 
reconstructed parent particle momentum with respect to the line of flight
less than 0.3 GeV, and (b) the momentum of the $\pi^+$ is greater 
than 10 GeV.  The shaded area in $\Xi^-\pi^+\pi^+$ and 
$\Sigma^+K^-\pi^+$ is the
estimated reflection from $\Lambda_{c}^+ \rightarrow \Sigma^-\pi^+\pi^+$
and $\Lambda_{c}^+ \rightarrow \Sigma^+K^-\pi^+$, respectively.  
The shape is determined by a Monte Carlo simulation and the area is 
normalized to the observed number of signal events in the $\Lambda_c^+$ data.

To estimate the total acceptance for decay modes of interest, we 
embedded Monte Carlo charm decays events in data events.
We generated charm events with an average transverse momentum 
$\langle p_{T} \rangle = 1.0$ GeV and longitudinal momentum spectrum as 
observed for the $\Xi_{c}^+$ data.  Detector hits, including
resolution and multiple Coulomb scattering smearing effects, produced
by these embedded tracks were OR'd into the hit banks of the underlying data 
event.  The new ensemble of hits was passed through the SELEX off-line 
reconstruction.  The total acceptance of each mode was determined from 
the number of fitted signal events from the charmed particle mass spectrum.

%% ONLY POSTSCRIPT FILES CAN BE INCLUDED

To check  the acceptance and particle identification, we measured two 
well--measured relative decay fractions,  
$B(D^0 \rightarrow K^-\pi^+\pi^+\pi^-)/
 B(D^0 \rightarrow K^-\pi^+) = 2.02 \pm 0.10 \pm 0.03$ and 
$B(D^{+} \rightarrow K^-K^+\pi^+)/
 B(D^{+} \rightarrow K^-\pi^+\pi^+) = 0.101 \pm 0.014 \pm 0.003$
where the first error is statistical and the second is the systematic 
difference between charge--conjugate states.  The results agree well with the 
world averages~\cite{PDG}.  For the $D^{+}$ decays, we applied a tighter cut
$L/\sigma$ ($>12$) to suppress background from 
$D_{s}^{+} \rightarrow K^-K^+\pi^+$ since the lifetime of $D^{+}$ is 2.4 times
longer than that of $D_{s}^{+}$.

To check the hyperon decay acceptance, we take into account two 
Cabibbo-favored modes
$\Xi_{c}^{+} \rightarrow \Sigma^{+}K^{-}\pi^{+}$ and 
$\Xi_{c}^{+} \rightarrow \Xi^{-}\pi^{+}\pi^{+}$. We measure the relative 
branching fraction to be $(1.12 \pm 0.35)$.  
The error is statistical only.
This result is comparable to the CLEO 
measurement, $(1.18 \pm 0.26 \pm 0.17)$~\cite{CLEO-3}.  
The number of events and estimated acceptance for these three $\Xi_{c}^+$ 
modes with the same set of cuts are summarized in table~\ref{tab xc+}.

\begin{table}
\caption{Summary of observed events and estimated acceptance 
 for $\Xi_{c}^+$ modes.}
\begin{tabular}{lrr} % c = centered and d = decimal justification.
                                        & Acceptance(\%)   &      Events \\
\tableline
$\Xi_{c}^+ \rightarrow \Xi^-\pi^+\pi^+$ & $1.61 \pm 0.04$  & $127 \pm 24$ \\
$\Xi_{c}^+ \rightarrow \Sigma^+K^-\pi^+$& $0.59 \pm 0.02$  & $ 52 \pm 13$ \\
$\Xi_{c}^+ \rightarrow pK^-\pi^+$       & $4.69 \pm 0.06$  & $ 85 \pm 22$ 
\end{tabular}
\label{tab xc+}
\end{table}

The preliminary branching fraction of the Cabibbo--suppressed decay
$\Xi_{c}^{+} \rightarrow pK^{-}\pi^{+}$ relative to the Cabibbo--favored
$\Xi_{c}^{+} \rightarrow \Sigma^{+}K^{-}\pi^{+}$ is measured to be 
$0.21 \pm 0.07$ which corresponds to 
$(4.0 \pm 1.3) \times {\rm tan}^{2}\theta_{c}$. 

\section{Conclusion}

We observed a large production asymmetry in favour of $\Lambda_c^+$ 
over $\overline{\Lambda}_{c}^{-}$ for all three beams in the 
region $x_F \ge 0.3$
The asymmetry is stronger for baryon beams than for the $\pi^-$ beam.   We
report the observation of the Cabibbo--suppressed decay 
$\Xi_{c}^{+} \rightarrow pK^{-}\pi^{+}$ at mass = ($2.467 \pm 0.001$) GeV
with $162 \pm 31$ signal events.  
The relative branching fraction of the decay
$\Xi_{c}^{+} \rightarrow pK^{-}\pi^{+}$ to the Cabibbo-favored
$\Xi_{c}^{+} \rightarrow \Sigma^{+}K^{-}\pi^{+}$ is measured to be
$B(\Xi_{c}^{+} \rightarrow pK^{-}\pi^{+}) /B(\Xi_{c}^{+} \rightarrow 
\Sigma^{+}K^{-}\pi^{+}) 
= 0.21 \pm 0.07$.

%%%%%%%%%%%%%%%%%%

\end{document}